\documentclass[12pt]{article}
\usepackage{epsfig}
\begin{document}
\begin{titlepage}
\begin{flushright}
\today\\
MSUHEP-060124\\
hep-ph/0602211
\end{flushright}
\vspace*{0.2cm}
\begin{center} \begin{large}
{\bf Higgs Boson Production and Decay in Little Higgs Models with T-parity}
\end{large} \end{center}

\begin{center}
\vspace*{0.2cm}
Chuan-Ren Chen, Kazuhiro Tobe and C.-P. Yuan\\
\vspace*{0.2cm}
Department of Physics and Astronomy\\
Michigan State University\\
East Lansing, MI 48824, USA\\
\end{center}

\vspace*{1cm}

\begin{abstract}
We study Higgs boson production and decay in a certain class of
Little Higgs models with T-parity in which some T-parity partners
of the Standard Model (SM) fermions gain their masses through
Yukawa-type couplings.
We find 
that the Higgs boson production cross section of a 120 GeV Higgs 
boson at the CERN LHC
via $gg$ fusion process at one-loop level
could be reduced by about $45\%$, $35\%$
and $20\%$, as compared to its SM prediction,
for a relatively low new particle mass scale $f=600,~700$ and $1000$ GeV, 
respectively.
On the other hand, the weak boson
fusion cross section is close to the SM value. Furthermore, the Higgs
boson decay branching ratio into di-photon mode can be enhanced by about
$35\%$ in small Higgs mass region in certain case, 
for the total decay width of Higgs boson in the Little Higgs
model is always smaller than that in the SM.
\end{abstract}
\end{titlepage}

\newpage
\section{Introduction}
In spite of the success of the Standard Model (SM) in describing
all the existing experimental data, the Higgs boson, which is
responsible for the electroweak symmetry breaking mechanism in the
SM, is yet-to-be found at the current Fermilab Tevatron Run II and
CERN Large Hadron Collider (LHC) experiments. The elementary Higgs
boson in the SM has a difficulty dubbed as ``naturalness problem'', i.e.,
its mass parameter can receive huge radiative corrections so
that it is difficult to understand why its mass is at the
electroweak scale, in contrast to the more fundamental scale, say,
Planck scale. Solving the naturalness problem has been one of the
driving forces to consider physics models beyond the SM.
Supersymmetry~\cite{SUSY} is one of the attractive solutions to this
problem. All the quadratic divergences in the Higgs mass parameter
induced by the SM particles in loops are canceled by those
induced by the loops of superpartners of SM particles, and hence
the electroweak scale Higgs boson mass parameter is protected from
large radiative corrections. Little Higgs
mechanism~\cite{LittleHiggs} provides another interesting
solution. Especially, the large correction induced by top-quark is
canceled by that from the heavy fermionic partner of the
top-quark, which is originated from the structure of the
approximate global symmetry in the top-sector. Similarly, the
quadratic divergence in the gauge-boson loops is canceled by that
from the extra heavy gauge-bosons which are the partners of SM 
gauge bosons.

In order to cancel the large quadratic divergence of the Higgs
mass parameter, one in general has to introduce new interactions
with the Higgs boson. Therefore, studying the consequence of these new
Higgs boson interactions would be very important for
understanding if such a cancellation is really happening. The new
Higgs boson interaction most likely affects $gg\rightarrow h$
production process where the top-Higgs interaction at one-loop
level plays a crucial role. It was pointed out in
Ref.~\cite{Han:2003gf} that the heavy fermionic partner of
top-quark could modify the Higgs boson production rate via gluon-gluon ($gg$)
fusion process in Little Higgs models.
Although many Little Higgs models suffer from
strong constraints from precision electroweak 
measurements~\cite{Csaki:2002qg}, recent studies 
in Refs.~\cite{Cheng:2003ju,Cheng:2004yc,Low:2004xc,Hubisz:2004ft,
Hubisz:2005tx,Cheng:2005as}
have shown that Littlest Higgs models~\cite{Arkani-Hamed:2002qy} 
with T-parity have weaker constraints,\footnote{The constraints can also be relaxed by adjusting the fermion hypercharges slightly from their original assignments~\cite{Gregoire:2003kr}.}
and therefore the new particle scale
can be significantly smaller than 1 TeV.

In this letter, we point out that there are other sets of new
Higgs boson interactions in a certain class of Little Higgs models with
T-parity. They are needed to generate mass terms for fermionic
T-parity partners of the SM fermions, and they can significantly
modify the production and decay of Higgs boson at high energy
colliders such as the Tevatron and the LHC. The rest of the paper
is organized as follows. In Sec.~2, we briefly review the model we study 
here; a Littlest Higgs model with T-parity. 
In Sec.~3, we discuss the production
rate of Higgs boson via $gg$ fusion process predicted by
this model at one-loop level.
In Sec.~4, we discuss the other production and decay processes of
the Higgs boson in this model.

\section{A Littlest Higgs model with T-parity}
The Littlest Higgs  model is based on an $SU(5)/SO(5)$ non-linear
sigma model~\cite{Arkani-Hamed:2002qy}. A vacuum expectation value
(VEV) of an $SU(5)$ $5 \times 5 $ symmetric tensor field ($\Sigma_0$) breaks the
$SU(5)$ to $SO(5)$ at the scale $f$ with
\begin{eqnarray}
\Sigma_0 &=& \left(
\begin{array}{lcr}
 & & \bf{1}_{2\times2}\\
 & 1 & \\
 \bf{1}_{2\times2} & &
\end{array}
\right).
\end{eqnarray}
A subgroup $[SU(2)_1\times U(1)_1]\times [SU(2)_2\times U(1)_2]$
of the $SU(5)$ is gauged, and at the scale $f$ it is broken into
the SM electroweak symmetry $SU(2)_L \times U(1)_Y$. The 14
Goldstone bosons $\Pi^a$ associated with this symmetry breaking
decompose under $SU(2)_L \times U(1)_Y$ as ${\bf 1_0}\oplus{\bf
3_{0}}\oplus {\bf 2_{1/2}}\oplus{\bf 3_{\pm 1}}$ and they are
parametrized by the non-linear sigma model field $\Sigma =\xi^2
\Sigma_0$ as the fluctuations around the VEV in the broken
directions, where $\xi=e^{i\Pi^a X^a/f}$ and $X^a$ are the broken
generators.
The $SU(2)$ doublet in the Nambu-Goldstone multiplet $\Pi^a$ is considered
to be the Higgs doublet~\cite{Arkani-Hamed:2002qy}. 

One way to ensure that the $\rho$-parameter
does not deviate from one  at tree level (similar to the SM) is to
introduce T-parity into the Little Higgs 
model~\cite{Cheng:2003ju,Cheng:2004yc,Low:2004xc}. Under T-parity,
the $[SU(2)_1\times U(1)_1]$ gauge fields transform to the
$[SU(2)_2\times U(1)_2]$ ones, and vice versa. 
The SM gauge fields are defined as the T-even combination of 
these gauge fields and the heavy extra gauge bosons are odd under T-parity.
Since T-parity is an exact symmetry in this model, the SM gauge bosons do not mix with 
the T-odd heavy gauge bosons, and consequently, the electroweak precision 
observables are not modified at tree
level. Beyond tree level, small
radiative corrections induced by the model to precision data still
allow the scale $f$ to be significantly lower than 1
TeV~\cite{Cheng:2004yc,Hubisz:2005tx}. Moreover, the Little Higgs
model with T-parity also provides a possible candidate for dark
matter~\cite{Cheng:2003ju,Cheng:2004yc,Hubisz:2004ft,Asano:2006nr}.

With T-parity in the Little Higgs models, the SM fermions are
T-even and their partners can in general be T-even and/or T-odd.
Since we have not seen such kind of
T-parity partners, especially T-odd fermions, we need to make them
heavy in order to be consistent with experimental measurements.
Here, as an interesting example, we consider a Littlest Higgs
model with T-parity proposed in
Refs.~\cite{Low:2004xc,Hubisz:2004ft,Hubisz:2005tx} (LH model). 
(See also the
original Littlest Higgs model with T-parity in
Ref.~\cite{Cheng:2004yc} and its variation in
Ref.~\cite{Cheng:2005as}.)

\subsection{Mass terms for T-odd fermions}
In the LH model~\cite{
Low:2004xc,Hubisz:2004ft,Hubisz:2005tx},
two fermion $SU(2)$ doublets $q_1$ and $q_2$ are introduced in such
a way that $q_A~(A=1,2)$ is transformed as a doublet
under $SU(2)_A$, and T-parity interchanges 
these two doublets.
The SM $SU(2)$ doublet is taken to be a T-even
combination of these doublets, and the T-odd
combination has to gain a heavy mass. In this section, let us briefly review
the mass terms for the T-odd fermions~\cite{
Low:2004xc,Hubisz:2004ft,Hubisz:2005tx}.

The fermion $SU(2)$ doublets $q_1$ and $q_2$ are embedded into
incomplete $SU(5)$ multiplets $\Psi_1$ and $\Psi_2$ as
$\Psi_1=(q_1,0,0_2)^{\rm T}$ and $\Psi_2=(0_2,0,q_2)^{\rm T}$,
where $0_2=(0,0)^{\rm T}$, and the doublets $q_1$ and $q_2$
are explicitly written as
$q_A=-\sigma_2 \left(
u_{L_A},
d_{L_A}
\right)^{\rm T}=\left(
i d_{LA},
-i u_{LA}
\right)^{\rm T}$
with $A=1,2$. (The superscript T denotes taking transpose.)
Under a global $SU(5)$ transformation,
the multiplets $\Psi_1$ and $\Psi_2$ transform as
$\Psi_1 \rightarrow V^* \Psi_1$ and $\Psi_2 \rightarrow V \Psi_2$,
where $V$ is an $SU(5)$ rotation matrix.
A multiplet $\Psi_c$ is also introduced as 
$\Psi_c=(q_c,\chi_c,\tilde{q}_c)^{\rm T}$,
whose transformation under the $SU(5)$ is non-linear:
$\Psi_c \rightarrow U \Psi_c$.
Here $U$ is the unbroken $SO(5)$ rotation and is a non-linear representation
of the $SU(5)$.
The object $\xi$ and the non-linear sigma model field
$\Sigma~(\equiv \xi^2 \Sigma_0)$
transform like
$\xi \rightarrow V\xi U^\dagger=U\xi \Sigma_0 V^T \Sigma_0$
and $\Sigma \rightarrow V\Sigma V^{\rm T}$, respectively,
under $SU(5)$. T-parity is implemented in fermion sector as well as
bosonic sector
and the transformation laws are defined as follows:
$\Psi_1 \leftrightarrow -\Sigma_0 \Psi_2$, 
$\Psi_c \rightarrow -\Psi_c$, and
$\xi \rightarrow \Omega \xi^\dagger \Omega$.
Thus,
$q_1\leftrightarrow -q_2$  and
$\Sigma \rightarrow \tilde{\Sigma}\equiv
\Sigma_0 \Omega \Sigma^\dagger \Omega \Sigma_0$
under T-parity. 
Here $\Omega \equiv {\rm diag}(1,1,-1,1,1)$.
Consequently, the Higgs boson is even under T-parity, and 
the T-invariant Lagrangian
for the mass terms of T-odd fermions can be written
as follows:
\begin{eqnarray}
{\cal L}_{\kappa}&=& -\kappa f (\bar{\Psi}_2 \xi \Psi_c
+\bar{\Psi}_1 \Sigma_0 \Omega \xi^\dagger \Omega\Psi_c)+{\rm h.c.}.
\label{kappa}
\end{eqnarray}
Note that $\xi$ contains the Higgs boson $h$, and therefore,
this Lagrangian
induces new Higgs boson interactions as well as the mass terms for
the T-odd fermions:
\begin{eqnarray}
{\cal L}_{\kappa}
&\simeq&-\sqrt{2} \kappa f
\left[
\bar{d}_{L_-} \tilde{d}_{c}+\frac{1+c_\xi}{2} \bar{u}_{L_-}
\tilde{u}_c
-\frac{s_\xi}{\sqrt{2}}\bar{u}_{L_-} \chi_c
-\frac{1-c_\xi}{2} \bar{u}_{L_-} u_c
 \right]+{\rm h.c.}\cdots.
\label{Kappa_int}
 \end{eqnarray}
Here we only showed the fermion mass terms and a few interaction terms
with the neutral Higgs boson.
$c_\xi (\equiv \cos\frac{v+h}{\sqrt{2}f})$ and
$s_\xi (\equiv \sin\frac{v+h}{\sqrt{2}f})$ are originated from 
the non-linear
sigma model field $\xi$, where $h$ and $v$ are
neutral Higgs boson field and its VEV, respectively. $u_{L_-}$
and $u_{L_+}$ are T-odd and T-even eigenstates,
respectively, as defined by $u_{L_\pm} = (u_{L_1}\mp
u_{L_2})/\sqrt{2}$. The same definition also applies to the $d$-quark.
Moreover, 
$\tilde{q}_c=(i\tilde{d}_c,-i\tilde{u}_c)^{\rm T}$,
and the second
component of the doublet $q_c$ is $-iu_c$. Note
that the T-odd combination of $q_1$ and $q_2$ 
together with $\tilde{q}_c$ 
gets a Dirac mass
$\sqrt{2} \kappa f$ in a limit of $v\rightarrow
0$, as shown in Eq.~(\ref{Kappa_int}). 
We also assume that the Dirac mass terms
($-m_q \bar{q}'_c q_c-m_\chi \bar{\chi}'_c \chi_c$)
make the remaining T-odd states heavy, and the extra
fields $q_c'$ and $\chi'_c$ are embedded into a complete $SO(5)$
multiplet, as suggested 
in Refs.~\cite{Cheng:2004yc,Low:2004xc,Hubisz:2004ft}, to avoid
inducing new quadratic divergence from radiative corrections.
The origin of these mass terms may come from new physics 
above the cutoff scale $4 \pi f$. In our analysis, 
we simply assume that the masses of these extra particles are
larger than the scale $f$. 
As to be shown later, it turns out that 
the radiative correction induced by these extra particles 
to the production
cross section of Higgs boson via $gg$ fusion, 
$\sigma_{gg  \to h}$, is not sensitive to
the actual values of these masses, as long as they are much larger than
half of the Higgs boson mass. 
We stress that the Higgs boson interactions
in ${\cal L}_{\kappa}$ provide $O(1)$ Yukawa-type interactions
for $\kappa \sim O(1)$, 
so that these individual interactions could contribute to the quadratic
divergences of Higgs boson mass. However, all the 
quadratic divergences, induced by the individual field 
introduced in Eq.~(\ref{kappa}), 
cancel in the limit of $v \rightarrow 0$, 
as long as $\Psi_c$ forms a complete $SO(5)$ multiplet. Hence,
the set of Higgs boson interactions introduced in ${\cal L}_{\kappa}$ is
consistent with the absence of large quadratic divergences
to the Higgs mass parameter. 
Before we examine the effect of these interactions to 
the production of Higgs boson via $gg$ fusion, we
first review the top Yukawa interactions in the Little Higgs model 
with T-parity.

\subsection{Top Yukawa interaction}

The large top Yukawa coupling 
generates a quadratic divergence to the Higgs boson mass.
In order to cancel the divergence, one introduces singlet fields
$U_1$ and $U_2$, which are embedded into the doublets:
$Q_1=(q_1,U_1,0_2)^{\rm T}$ and $Q_2=(0_2,U_2,q_2)^{\rm T}$,
and constructs the following T-parity invariant
Lagrangian~\cite{Low:2004xc,Hubisz:2004ft,Hubisz:2005tx}:
\begin{equation}
{\cal L}_t= -\frac{\lambda_1}{2\sqrt{2}}f\epsilon_{ijk} \epsilon_{xy}
\left[(\bar{Q}_1)_i \Sigma_{jx} \Sigma_{ky}-
(\bar{Q}_2 \Sigma_0)_i \tilde{\Sigma}_{jx} \tilde{\Sigma}_{ky}
\right] u_R
-\lambda_2 f (\bar{U}_1 U_{R_1}+\bar{U}_2 U_{R_2}) +{\rm h.c.},
\label{top_yukawa_int}
\end{equation}
where $\epsilon_{ijk}$ and $\epsilon_{xy}$ are antisymmetric tensors,
and $i,~j$ and $k$ run over $1-3$ and $x$ and $y$ over $4-5$.
Under T-parity, these fields transform as
$Q_1 \leftrightarrow -\Sigma_0 Q_2~$,
$U_{R_1}\leftrightarrow -U_{R_2}$ and $u_R\rightarrow u_R$.
The above Lagrangian contains the following neutral Higgs boson 
interactions:
\begin{eqnarray}
{\cal L}_t &\simeq&
-\lambda_1 f \left(
\frac{s_\Sigma}{\sqrt{2}} \bar{u}_{L_+} u_R
+\frac{1+c_\Sigma}{2} \bar{U}_{L_+} u_R
\right)
-\lambda_2 f \left(\bar{U}_{L_+} U_{R_+}+\bar{U}_{L_-} U_{R_-}
\right)+{\rm h.c.}\, ,
\label{top-yukawa}
\end{eqnarray}
where $c_\Sigma(\equiv\cos\frac{\sqrt{2}(v+h)}{f})$ and
$s_\Sigma(\equiv \sin\frac{\sqrt{2}(v+h)}{f})$
are originated from the non-linear sigma model field $\Sigma$.
We have defined T-parity eigenstates as
$U_{L_\pm}=\frac{U_1\mp U_2}{\sqrt{2}}$
and $U_{R_\pm}=\frac{U_{R_1}\mp U_{R_2}}{\sqrt{2}}$.
One T-odd Dirac fermion $T'$ ($T'_L \equiv U_{L_-},~T'_R \equiv U_{R_-}$)
gets a mass $m_{T'}=\lambda_2 f$ (cf. Eq.~(\ref{top-yukawa})), and 
a T-odd combination of the doublets $q_1$ and $q_2$
obtains a mass from ${\cal L}_\kappa$ (cf. Eq.~(\ref{Kappa_int})).
Note that $T'$ does not have tree level Higgs boson interaction, and thus
it does not contribute to the $gg$ fusion process at the one-loop order.
The left-handed (or right-handed) top quark is a linear combination of
$u_{L_+}$ and $U_{L_+}$ (or $u_R$ and $U_{R_+}$) and another independent
linear combination is a heavy T-even partner $(T)$ of top quark with the mass
$m_{T}\simeq \sqrt{\lambda_1^2+\lambda_2^2} f$ which is responsible for 
canceling the quadratic divergence to the Higgs mass induced by the top quark.

From Eq.~(\ref{top-yukawa}), we find that the Higgs boson interactions
are approximately given by
\begin{eqnarray}
-{\cal L} &=& g_{h\bar{t}t} h \bar{t}_L t_R
+g_{h\bar{T}T}h \bar{T}_L T_R+{\rm h.c.},
\end{eqnarray}
where
\begin{eqnarray}
g_{h\bar{t}t} &\simeq &\frac{m_t}{v_{SM}}  \left\{
1-\frac{3+2R^2+3R^4}{4(1+R^2)^2}
\frac{v_{SM}^2}{f^2}+\cdots
\right\},~~
g_{h\bar{T}T} \simeq
-\frac{m_t}{v_{SM}}\frac{R}{
1+R^2}\frac{v_{SM}}{f}+\cdots,
\label{top_yukawa}
\end{eqnarray}
with $R=\lambda_1/\lambda_2$ and $v_{SM}\equiv
f\sqrt{1-\cos\frac{\sqrt{2}v}{f}}$. With this definition of
$v_{SM}$, gauge boson masses and Fermi constant are expressed,
similar to those in the SM, by
$m_Z=\sqrt{g^2+g^{'2}}v_{SM}/2$, $m_W=g v_{SM}/2$, and
$G_F=1/(\sqrt{2}v^2_{SM})$ at tree level, respectively. Hence,
$v_{SM}$ is equivalent to the Higgs boson VEV in the SM;
$v_{SM}\simeq 246$ GeV. It is important to note that the relation
between the top mass and its Yukawa coupling is modified in this
model; the top Yukawa coupling is reduced, compared to that in the
SM. In addition, the heavy $T$-quark also has Yukawa
interaction, but its sign is opposite to that of the top Yukawa
coupling. The modification of top-Yukawa coupling and the new
Yukawa interaction of $T$-quark will be important for 
studying the Higgs boson production rate via $gg$ fusion process and
the decay branching ratio of Higgs boson into di-photon mode.

\subsection{Other Higgs boson interactions}
Here we summarize other Higgs interactions that are important
when we consider Higgs boson productions and decays.

Yukawa couplings of up-type quarks for the first and second generations
are given by the similar Lagrangian for the top quark
(cf. Eq.~(\ref{top_yukawa_int})), but without introducing extra singlet fields
like $U_i$ and $U_{R_i}$ $(i=1-2)$ in Eq.~(\ref{top_yukawa_int}).
For down-type quark Yukawa couplings, one of the possible effective
Lagrangians~\cite{Hubisz} is given by
\begin{eqnarray}
{\cal L}_{\rm down} &=&\frac{i\lambda_d}{2\sqrt{2}}
f \epsilon_{ij} \epsilon_{xyz}\left[
(\bar{\Psi}'_2)_x \Sigma_{i y} \Sigma_{j z} X
-(\bar{\Psi}'_1 \Sigma_0)_x \tilde{\Sigma}_{i y}
\tilde{\Sigma}_{j z} {\tilde{X}}
\right]d_{R},
\label{down-yukawa}
\end{eqnarray}
where $\Psi'_1=(-\sigma_2 q_1,0,0_2)^{\rm T}$ and 
$\Psi'_2=(0_2,0,-\sigma_2 q_2)^{\rm T}$.
Here $X$ transforms into $\tilde{X}$ under T-parity, and 
it is a singlet under $SU(2)_i~(i=1-2)$
and its $U(1)_i~(i=1-2)$ charges are $(Y_1,~Y_2)=(1/10,~-1/10)$.
In this paper, we consider two possible choices\footnote{
We thank Jay Hubisz
for pointing out these possibilities to us.
The corresponding Lagrangian proposed in Ref.~\cite{Hubisz:2004ft},
cf. Eq.~(2.35), is not invariant under $U(1)_i~(i=1,2)$.
}
for $X$:
$X=(\Sigma_{33})^{-1/4}$ (denoted as Case A) and 
$X=(\Sigma^\dagger_{33})^{1/4}$
(denoted as Case B), where $\Sigma_{33}$ is the 
$(3,3)$ component of the non-linear
sigma model field $\Sigma$.

An important point we note is that all the Yukawa couplings 
$g_{h \bar{F} F}$ (where $F$ denotes a SM fermion)
are modified from those in the SM 
($g_{h \bar{F} F}^{\rm SM}$).
Their ratios are approximately given as follows:
\begin{eqnarray}
\frac{g_{h\bar{u}u}}{g_{h\bar{u}u}^{\rm SM}} 
&=& 1
-\frac{3}{4}\frac{v_{SM}^2}{f^2}-\frac{5}{32}\frac{v_{SM}^4}{f^4}+\cdots
\simeq \left\{
\begin{array}{l}
0.90~~{\rm for}~f=700~{\rm GeV},\\
0.95~~{\rm for}~f=1~{\rm TeV},
\end{array}
\right.
\label{Higgs-up}
\end{eqnarray}
for $u=$ up and charm quarks (see Eq.~(\ref{top_yukawa}) 
for the top Yukawa coupling). For down-type quarks,
\begin{eqnarray}
\frac{g_{h\bar{d}d}}{g_{h\bar{d}d}^{\rm SM}} 
&=&1-\frac{1}{4}\frac{v_{SM}^2}{f^2}+\frac{7}{32}
\frac{v_{SM}^4}{f^4}+\cdots
\simeq \left\{
\begin{array}{l}
0.97~~{\rm for}~f=700~{\rm GeV},\\
0.99~~{\rm for}~f=1~{\rm TeV},
\end{array}
\right.~~{\rm for~Case~A},
\label{Higgs-downA}
\\
&=&1-\frac{5}{4}\frac{v_{SM}^2}{f^2}-\frac{17}{32}
\frac{v_{SM}^4}{f^4}+\cdots
\simeq \left\{
\begin{array}{l}
0.84~~{\rm for}~f=700~{\rm GeV},\\
0.92~~{\rm for}~f=1~{\rm TeV},
\end{array}
\right.~~{\rm for~Case~B}.
\label{Higgs-downB}
\end{eqnarray}
Note that the down-type quark Yukawa couplings could be
significantly suppressed in Case B.
We consider the same Yukawa structures in lepton sector,
as in quark sector. Thus, the charged lepton Yukawa couplings
are also suppressed in the same way as the down-type Yukawa
couplings.
Similarly, Higgs boson interactions with SM gauge bosons 
$g_{hVV}$ (where $V=Z,W$) are
also slightly suppressed:
\begin{eqnarray}
\frac{g_{hVV}}{g_{hVV}^{\rm SM}} &=& 1-
\frac{1}{4}\frac{v^2_{SM}}{f^2}-\frac{1}{32}\frac{v^4_{SM}}{f^4}
\cdots
\simeq  \left\{
\begin{array}{l}
0.97~~{\rm for}~f=700~{\rm GeV},\\
0.98~~{\rm for}~f=1~{\rm TeV}.
\end{array}
\label{Higgs-gauge}
\right.
\end{eqnarray}

In addition to the Higgs boson interactions with the SM particles, 
there are Higgs couplings with heavy extra-gauge bosons and
triplet Higgs bosons, as shown in Ref.~\cite{Han:2003gf}.
All the above mentioned Higgs boson interactions are important
for Higgs boson productions and decays as we will discuss later.

\section{Higgs boson production via $gg$ fusion process}
\subsection{Contributions from T-odd fermions}
\begin{figure}[ht]
\centering
\includegraphics*[width=13cm,angle=0]{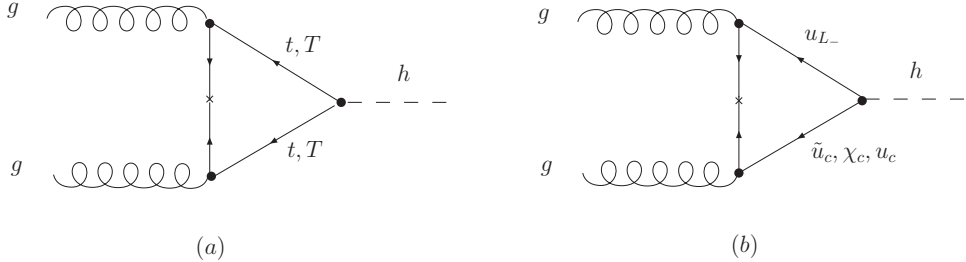}
\caption{Contributions to Higgs boson production via $gg$ fusion process
$gg \rightarrow h$,
induced by (a) top-quark and T-even partner $T$, and (b) T-odd fermions.}
\label{ggh}
\end{figure}
Due to the new Higgs boson interaction in ${\cal L}_{\kappa}$ of
Eq.~(\ref{Kappa_int}),
the T-odd fermions can contribute to the Higgs boson production via
$gg$ fusion process at one-loop level,
as shown in Fig.~{\ref{ggh}} (b).
From the structure of the mass matrix and Higgs boson interactions
for the T-odd fermions, we find that
 $h \bar{u}_{L_-}\tilde{u}_c$ interaction provides the dominant
T-odd fermion contribution to $\sigma_{gg\rightarrow h}$ and the result
is not sensitive to the masses $m_q$ and $m_\chi$, as long as the 
T-odd fermion masses are much larger than half of the Higgs
boson mass.
The ratio of the amplitude induced by the T-odd fermions to the 
one by the SM top-quark, which is the dominant contribution
in the SM, is approximately expressed as
\begin{eqnarray}
\frac{A_{gg\rightarrow h}({\rm T\mbox{-}odd~ fermion})}
{A_{gg \rightarrow h}({\rm top~in~SM})} &\simeq&
-\frac{1}{4}\frac{v_{SM}^2}{f^2} +\cdots
\simeq \left\{
\begin{array}{l}
-3\% ~~{\rm for}~f=700~{\rm GeV},\\
-1.5\%~~{\rm for}~f=1~{\rm TeV}.
\end{array}
\right.
\label{result}
\end{eqnarray}
Here, we have assumed that
the fermions in the loop are much heavier than half of the Higgs boson,
so that the one-loop vertex diagram in Fig.~\ref{ggh} can be
approximated as a three-point vertex after shrinking the heavy internal
lines into a point~\cite{Gunion:1989we}.
The negative sign of the ratio is originated from the positive
sign of $h \bar{u}_{L_-} \tilde{u}_c$ interaction term in Eq.~(\ref{Kappa_int})
after fixing the correct negative 
sign for the $\bar{u}_{L_-} \tilde{u}_c$-mass term.
Note that as shown in Eq.~(\ref{result}) the leading order contribution,
in terms of $v_{SM}/f$,
does not explicitly depend on the parameter $\kappa$. 
Namely,  $\kappa$ term generates a ``non-decoupling''
contribution to $\sigma_{gg \to h}$ which does not vanish as 
$\kappa f \to \infty$ with a fixed $f$ value.
Since the interaction shown in Eq.~(\ref{kappa}) is needed for each
fermion generation
to generate mass terms for all the T-odd partners,
the parameter $\kappa$ has a generation index in general~\cite{Hubisz:2005bd}.
Since the result in
Eq.~(\ref{result}) does not depend on $\kappa$, the sum over all three
generations of this type of corrections to the $gg\rightarrow h$
amplitude will be
three times of the result shown in Eq.~(\ref{result}).
(As shown in Eq.~(\ref{Kappa_int}), there are no equivalent Higgs couplings
to down-type quarks in ${\cal L}_\kappa$.)
Hence, the correction $\delta \sigma_{gg \rightarrow h}$
to the production cross section of
$gg \rightarrow h$ induced by the T-odd fermions
is approximately given by
\begin{eqnarray}
\frac{\delta \sigma_{gg \rightarrow h}( {\rm T\mbox{-}odd
~ fermions})}
{\sigma_{gg \rightarrow h}({\rm top~in~SM})}
&\simeq& -\frac{3}{2}\frac{v_{SM}^2}{f^2} 
+\cdots
\simeq
\left\{
\begin{array}{l}
-19\%~~{\rm for}~~f=700~{\rm GeV},\\
-9\% ~~{\rm for}~~f=1~{\rm TeV},
\end{array}
\right.
\label{T_odd_result}
\end{eqnarray}
for three generation case.
Therefore, we find that the effect of the T-odd fermion mass terms 
on the Higgs boson production rate via $gg$ fusion could be significant, 
especially when $f$ is below 1 TeV.

\subsection{Contributions from top and heavy T-even top partner}
In the SM, the most important contribution to the Higgs boson 
production
via $gg$ fusion process comes from top-quark loop, as shown in 
Fig.~\ref{ggh} (a).
As we have discussed, the top-Yukawa coupling is modified in the LH
model,
and hence the contribution to the $gg$ fusion process is also modified.
Furthermore, there is 
also new contribution induced at one-loop level 
by the partner of top-quark, the
T-even heavy quark $T$.
The ratios of the amplitudes to the top contribution in the SM
are given as follows:
\begin{eqnarray}
\frac{A_{gg\rightarrow h}(\rm{top~in~LH})}
{A_{gg\rightarrow h}(\rm{top~in~SM})}
&\simeq& 1-\frac{3+2R^2+3R^4}{4(1+R^2)^2}\frac{v_{SM}^2}{f^2}
+\cdots.\\
\frac{A_{gg_\rightarrow h}(T~{\rm in~LH})}
{A_{gg \rightarrow h}({\rm top~in~SM})}
&\simeq& -\frac{m_t^2}{m_T^2}+\cdots =-\frac{R^2}{(1+R^2)^2}
\frac{v_{SM}^2}{f^2}+\cdots,
\end{eqnarray}
where we have assumed $m_t,~m_T\gg m_h/2$.
Therefore, the cross section of the Higgs boson production
via $gg$ fusion in the LH model is modified by the T-even top sector
(including both top and $T$ contributions). 
As compared to that in the SM,
\begin{eqnarray}
\frac{\delta \sigma_{gg \rightarrow h}({\rm T}\mbox{-}{\rm even~top~sector})}
{\sigma_{gg \rightarrow h}({\rm top~in~SM})}
&\simeq&
-\frac{3}{2} \frac{v_{SM}^2}{f^2}+\cdots
\simeq
\left\{
\begin{array}{l}
-19\%~~{\rm for~}f=700~{\rm GeV},\\
-9\%~~{\rm for~}f=1~{\rm TeV}.
\end{array}
\right.
\label{T-even_result}
\end{eqnarray}
We note that although top and $T$ contributions separately depend on
$R$, the sum of them does not.
This suggests that even if $\lambda_2$ is large, and therefore
$T$-quark is heavy, the $T$ contribution does not decouple as long as
the scale $f$ is about 1 TeV.
In case of the Littlest Higgs model without T-parity,
the authors in Ref.~\cite{Han:2003gf} had reached a similar conclusion
on the contribution from top and heavy T-even
top partner which is rather common in any Littlest Higgs models.\footnote{
Since in the Little Higgs model without T-parity, 
the $\rho$ parameter at tree level
is not one~\cite{Csaki:2002qg}, the model has stronger 
constraint on the scale $f$.
Therefore, one expects that the effect on the Higgs boson production via
$gg$ fusion in the model without T-parity 
will be much smaller than what is expected in the 
Littlest Higgs model with T-parity.}
Since the T-odd fermion contribution discussed
in the previous section is as large as the one induced
by the T-even top-quark sector, the correction to $\sigma_{gg\rightarrow h}$ 
in the LH model is largely enhanced 
by the T-odd fermion contributions
as compared to that in the Littlest Higgs model without T-parity.

When we sum over all the contributions discussed above, 
the deviation
($\delta\sigma_{gg\rightarrow h}\equiv\sigma_{gg\rightarrow h}^{\rm LH}
-\sigma_{gg\rightarrow h}^{\rm SM}$) of the Higgs boson 
production cross section
via $gg$ fusion process in the LH model
($\sigma_{gg\rightarrow h}^{\rm LH}$)
from the SM prediction ($\sigma_{gg\rightarrow h}^{\rm SM}$)
is approximately given by
\begin{eqnarray}
\frac{\delta \sigma_{gg\rightarrow h}}
{\sigma_{gg\rightarrow h}^{\rm SM}}
&\simeq& -3 \frac{v_{SM}^2}{f^2}+\cdots
\simeq \left\{
\begin{array}{l}
-37\% ~{\rm for} ~f=700~{\rm GeV},\\
-18\%~{\rm for} ~f=1~{\rm TeV},
\end{array}
\right.
\label{total}
\end{eqnarray}
where we have assumed that the Higgs mass is smaller
than the fermion masses in the loop. It is clear that the extra 
contributions in the
LH model significantly suppress the
Higgs boson production cross section via $gg$ fusion process.
\begin{figure}[t]
\centering
\includegraphics*[width=12cm,angle=0]{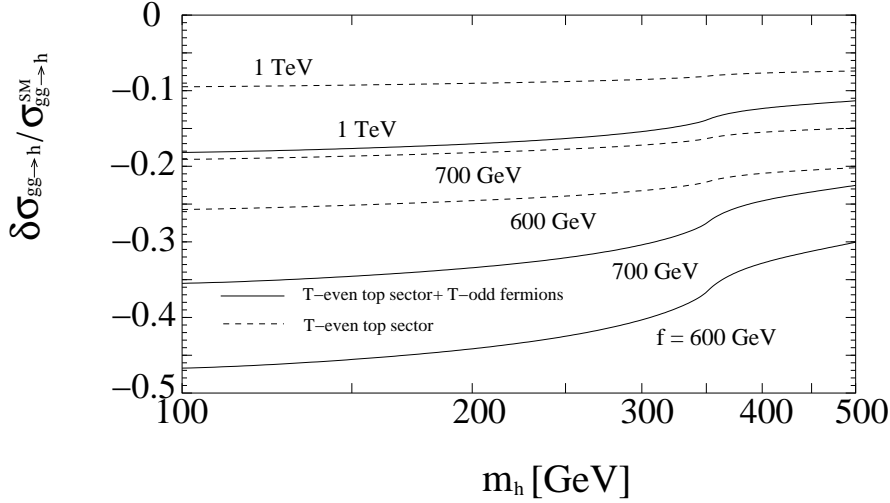}
\caption{Deviations
$(\delta \sigma_{gg\rightarrow h}=\sigma_{gg \rightarrow h}^{\rm LH}
-\sigma_{gg \rightarrow h}^{\rm SM})$
of the Higgs boson production cross section via
$gg$ fusion process in the LH model
$(\sigma_{gg \rightarrow h}^{\rm LH})$
from that in the SM $(\sigma_{gg \rightarrow h}^{\rm SM})$,
normalized by $\sigma_{gg \rightarrow h}^{\rm SM}$,
as a function of Higgs mass $m_h$.
We have taken $\kappa=3$, $m_q=m_\chi=5f$ and $R=1$, though
our result is not sensitive to their specific values as long as
$m_q,~m_\chi\gg m_h/2$.
Dashed lines show the effect
induced by the T-even top sector only.
Solid lines include the contributions from T-odd fermions
in addition to T-even top sector.
In each case, the results for $f=600$ GeV, $700$ GeV and $1$ TeV
are shown. Here, the complete one-loop calculation was used
in our numerical analysis.}
\label{d_sigma}
\end{figure}
In Fig.~\ref{d_sigma}, we show a numerical result of the deviation
$(\delta \sigma_{gg\rightarrow h})$ of the Higgs boson
production cross section via $gg\rightarrow h$ in the LH model 
from the SM prediction, normalized by the SM prediction
($\delta\sigma_{gg\rightarrow h}/\sigma_{gg \rightarrow h}^{\rm SM}$).
Here, we assumed that $\kappa=3$, $m_q=m_\chi=5f$ and $R=1$, 
but we have checked 
that our result does not strongly depend on these parameters
as long as $m_q$ and $m_\chi$ are much larger than $m_h/2$.
For our numerical analysis, we adapted a publically available code,
{\tt HDECAY}~\cite{Djouadi:1997yw}, 
for the SM calculation, and modified the code with a complete one-loop 
calculation
in accordance with the effective Lagrangian described
in the previous section for the LH model calculation.
Dashed lines show the corrections induced by the 
T-even top sector only
(assuming there are no other corrections), 
and solid lines include the contributions
from T-odd fermions 
in addition to T-even top sector.
One sees that the approximate results in
Eqs.~(\ref{T_odd_result}), (\ref{T-even_result}) and (\ref{total}) 
well describe the numerical result in
Fig.~\ref{d_sigma} when Higgs mass is small.
Fig.~\ref{d_sigma} show that, if the scale $f$ is smaller than
about 1 TeV, the production cross section
via $gg$ fusion is largely suppressed in all range of Higgs mass,
but especially in small Higgs mass region. For example,
the deviation from the SM prediction
can be more than $40\%~(30\%)$ for $m_h<300$ GeV and $f<600~(700)$ GeV
if we take into account all the corrections discussed above.
This large suppression will be
important especially when the Higgs mass is relatively small
because the $gg$ fusion process is one of the main discovery
modes for detecting a light SM Higgs boson.

We note that as suggested by the naive dimensional analysis~\cite{Manohar:1983md}
in low energy effective theory, 
one may write down the operator,
$\sum_{i=1,2}\frac{g_s^2 f^2}{\Lambda^2}\Sigma^\dagger_{3i}\Sigma_{3i} 
G_{\mu\nu}^A G^{A,\mu\nu}\sim \frac{g_s^2 h^\dagger h}{\Lambda^2} 
G_{\mu\nu}^A G^{A,\mu\nu}$, with an $O(1)$ coefficient, 
where $G^{A}_{\mu\nu}$ is the field strength tensor of
the gluon field and $g_s$ is the strong coupling. 
In that case, this operator will induce a counterterm for
$h G_{\mu\nu}^A G^{A,\mu\nu}$ coupling with its coefficient at the order of
$g_s^2v/(16\pi^2f^2)$, which has the same magnitude as the one-loop
contribution calculated above.
However, due to the Little Higgs mechanism, we expect the coefficient for
the operator $\Sigma^\dagger_{3i}\Sigma_{3i}$, which generates Higgs 
boson mass term, to be suppressed by a two-loop suppression 
factor of $v^2/\Lambda^2\simeq (1/16\pi^2)^2$
as compared to the naive dimensional analysis (with a coefficient 
$f^2\Lambda^2$). It is likely the same suppression factor $(1/16\pi^2)^2$ also applies
to the above operator $\Sigma^\dagger_{3i}\Sigma_{3i} 
G_{\mu\nu}^A G^{A,\mu\nu}$ and yields a much smaller coefficient
as compared to the genuine one-loop contributions.
Therefore, we expect the one-loop contributions discussed above
well represent the dominant contributions to $\sigma_{gg \rightarrow h}$.

\section{Other Higgs boson production channels and decay modes}

In the LH model, the Higgs boson interactions are modified through the
interactions of the non-linear sigma model field with other particles,
as shown in Eqs.~(\ref{top_yukawa}), (\ref{Higgs-up}), 
(\ref{Higgs-downA}), (\ref{Higgs-downB}) and (\ref{Higgs-gauge}).

Because of the slight suppression of the Higgs couplings with
gauge bosons and top quark,
the Higgs boson production rates via the gauge boson fusion processes
($VV\rightarrow h$ with $V=W,Z$),
the associated $Wh$ and $\bar{t}th$ processes
are also slightly suppressed.

\begin{figure}
\centering
\includegraphics*[width=16cm,angle=0]{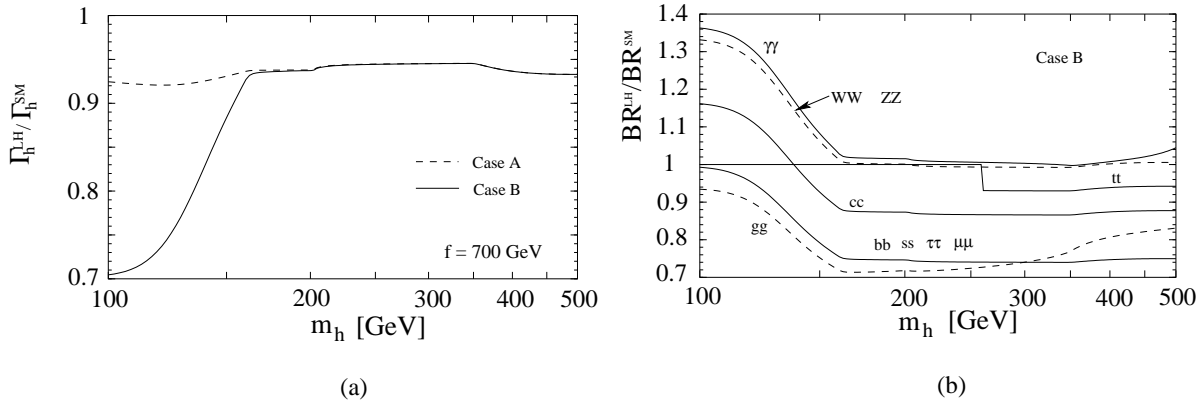}
\caption{(a) A ratio of the total Higgs decay width in the LH model
$\Gamma_h^{\rm LH}$ to one in the SM $\Gamma_h^{\rm SM}$ for
$f=700$ GeV in Case A and B for the down-type quark Yukawa couplings.
(b) Ratios of the Higgs decay branching ratios in the LH
model $\Gamma_h^{\rm LH}$ to those in the SM $\Gamma_h^{\rm SM}$ for
$f=700$ GeV in Case B.}
\label{decay_width}
\end{figure}

Furthermore, the modification of the Higgs couplings also affects
the decay branching ratios of Higgs boson.
In Fig.~\ref{decay_width} (a), we show the ratios of the total decay 
width of Higgs boson
in the LH model to that in the SM,
$\Gamma_h^{\rm LH}/\Gamma_h^{\rm SM}$, for $f=700$ GeV,
in Case A and B of the down-type quark Yukawa interactions, cf. 
Eq.~(\ref{down-yukawa}).
In Case A, partial decay widths for main decay modes of Higgs 
boson such as $b\bar{b}$, $\tau\tau$ and $VV~(V=W,Z)$ are almost
equally suppressed by about $(0.97)^2=0.94$ for $f=700$ GeV.
Therefore, the total decay width of Higgs boson is
almost uniformly suppressed in the whole range of Higgs mass.
Consequently, the branching ratios of most of the Higgs decay modes
are very close to the SM predictions. Thus, in Case A, the only
sizable change from the SM prediction is the $gg$ fusion
cross section $\sigma_{gg\rightarrow h}$, as discussed in the previous
section.

On the other hand, in Case B, both bottom and tau Yukawa couplings
are significantly suppressed, and hence the total decay width
of Higgs boson is largely reduced in the small Higgs mass
region where $h\rightarrow b\bar{b}$ and $\tau\tau$ decay
modes dominate, as seen in Fig.~\ref{decay_width} (a).
For $m_h$ larger than $2m_W$, the Higgs boson mostly decays into
gauge bosons, $ZZ$ and $WW$, and the suppression factor of 
the total decay width
is about $(0.97)^2=0.94$ for $f=700$ GeV. 

An interesting effect in Case B is that because of the largely reduced 
total decay width in small Higgs mass region, 
some of the Higgs boson decay branching ratios are increased even though 
the corresponding
partial decay widths are reduced. We have also calculated the 
di-photon decay width $\Gamma(h\rightarrow \gamma \gamma)$
at one-loop level.\footnote{See Ref.~\cite{Han:2003gf} for the study
of $h\rightarrow \gamma\gamma$ in the Littlest Higgs model without T-parity.}
Similar to the fact that the W-boson contribution dominates over the top-quark contribution
to $\Gamma(h\rightarrow \gamma \gamma)$
for small Higgs mass region in the SM, the effect of the
extra-fermions in the di-photon decay mode is less
important than that in the $gg$ fusion process. Furthermore,
the extra-boson contributions tend
to cancel the extra-fermion contributions in di-photon decay mode,
and therefore the partial decay width of $h\rightarrow \gamma \gamma$ does
not change very much, as compared to the SM prediction. As a result, 
in contrast to Case A, the decay
branching ratio of $h\rightarrow \gamma \gamma$ is enhanced by
about 35\% for a 100 GeV Higgs boson in Case B, for the total decay width of Higgs boson is 
reduced by about 30\%.

In Fig.~\ref{decay_width} (b), we show a numerical result on
ratios of Higgs decay branching ratio
in the LH model to that in the SM,
${\rm BR}^{\rm LH}/{\rm BR}^{\rm SM}$, for a few 
main Higgs decay modes in Case B. 
Here we have again taken $f$ to be $700$ GeV.
We see that especially $\gamma\gamma$ and $VV$ ($V=W,Z$) decay
branching ratios are largely enhanced in the small Higgs mass region.

\begin{table}[t]
\begin{center}
\begin{tabular}{|r||c|c|c|c|}
\hline
$m_h=120$ GeV & $R_{{\rm BR}(\gamma \gamma)}$ &$R_{{\rm BR}(\tau \tau)}$
& $R_{{\rm BR}(b \bar{b})}$ & $R_{{\rm BR}(VV)}$\\
\hline \hline
$R_{\sigma(gg)}$ (Case A)& {0.57,~0.68,~0.84} & {0.56,~0.67,~0.83}
&$-$&0.55,~0.66,~0.83\\
 (Case B)& {0.81,~0.86,~0.93} & {0.51,~0.63,~0.81}
&$-$&0.78,~0.84,~0.92\\
\hline
$R_{\sigma(VV)}$ (Case A)& {0.97,~0.98,~0.99} & {0.95,~0.96,~0.98}
&$-$&0.94,~0.96,~0.98 \\
 (Case B)& {1.34,~1.22,~1.09} & {0.84,~0.89,~0.95}
&$-$&1.30,~1.19,~1.08 \\
\hline
$R_{\sigma(t\bar{t}h)}$ (Case A)
&$ -$&{0.87,~0.90,~0.95} &{0.87,~0.90,~0.95}& $-$\\
(Case B) &$ -$&{0.77,~0.83,~0.92} &{0.77,~0.83,~0.92}& $-$\\
\hline
$R_{\sigma(Vh)}$ (Case A)
&{0.97,~0.98,~0.99} & $-$ & {0.95,~0.96,~0.98} & $-$\\
(Case B) &{1.34,~1.22,~1.09} & $-$ & {0.84,~0.89,~0.95} & $-$\\
\hline\hline
$m_h=200$ GeV & $R_{{\rm BR}(\gamma \gamma)}$ &$R_{{\rm BR}(\tau \tau)}$
& $R_{{\rm BR}(b \bar{b})}$ & $R_{{\rm BR}(VV)}$\\
\hline\hline
$R_{\sigma(gg)}$ (Case A)&$-$ & $-$ & $-$& {0.55,~0.67,~0.83}\\
(Case B)&$-$ & $-$ & $-$& {0.56,~0.67,~0.83}\\
\hline
$R_{\sigma(VV)}$ (Case A)&$-$ & $-$ & $-$& {0.90,~0.94,~0.97}\\
(Case B)&$-$ & $-$ & $-$& {0.90,~0.94,~0.97}\\
\hline
\end{tabular}
\end{center}
\label{discovery_pot}
\caption{$R_{\sigma}\times R_{\rm BR}$ for $f=(600,~700,~1000)$ GeV.
Here $R_{\sigma (X)}
(\equiv \sigma_{(X)}^{\rm LH}
/\sigma_{(X)}^{\rm SM}$) is
defined as a ratio of the Higgs boson production cross section in the
little Higgs model $(\sigma_{(X)}^{\rm LH})$ to one in the SM 
$(\sigma_{(X)}^{\rm SM})$ for each Higgs boson production process $X$.
The subscripts
$gg,~VV,~t\bar{t}h,~$ and $Vh$ represent $gg$ fusion ($gg\rightarrow h$),
weak boson fusion ($VV\rightarrow h$ where $V=W,Z$), 
$t\bar{t}h$ and $Vh$ associated productions, respectively.
$R_{{\rm BR}(Y)}
\equiv {\rm BR}_{(Y)}^{\rm LH}
/{\rm BR}_{(Y)}^{\rm SM}$
for each Higgs decay mode $h\rightarrow Y$, where $Y=\gamma\gamma,~\tau\tau,~
b\bar{b}$ and $VV$.}
\end{table}

Since the production cross section via $gg$ fusion is strongly suppressed,
the Higgs discovery modes will be changed significantly.
In Table 1, $R_{\sigma}\times R_{\rm BR}$
is listed for various production and decay processes in cases for 
$m_h=120$ GeV and $200$ GeV.
Here, we define $R_{\sigma (X)}(\equiv \sigma_{(X)}^{\rm LH}
/\sigma_{(X)}^{\rm SM})$
as the ratio of the Higgs boson production cross section in the LH
model $(\sigma_{(X)}^{\rm LH})$ to that in the SM
$(\sigma_{(X)}^{\rm SM})$
for each production process X. The subscripts $gg,~VV,~t\bar{t}h,~$ and $Vh$ 
represent $gg$ fusion ($gg\rightarrow h$),
weak boson fusion ($VV\rightarrow h$, with $V=W,Z$), 
$t\bar{t}h$ and $Vh$ associated productions, respectively. 
$R_{{\rm BR}(Y)}\equiv {\rm BR}_{(Y)}^{\rm LH}/
{\rm BR}_{(Y)}^{\rm SM}$
for $h\rightarrow Y$ decay modes, where $Y=\gamma\gamma,~\tau\tau,~
b\bar{b}$ and $VV$.

In the SM, the $\gamma \gamma$ decay mode of Higgs boson 
produced via $gg$ fusion
is one of the important discovery channels for a light
Higgs boson with mass around 100 GeV. However, in the
LH model that we study here, 
this mode would be strongly suppressed. For example,
$R_{\sigma(gg)}\times R_{{\rm BR}(\gamma \gamma)}=0.68~(0.86)$
for $m_h=120$ GeV and $f=700$ GeV in Case A (Case B), as shown
in Table 1. 
Note that since the $h\rightarrow \gamma\gamma$ decay branching
ratio is enhanced in Case B, 
$R_{\sigma(gg)}\times R_{{\rm BR}(\gamma \gamma)}$ is not suppressed
as largely as in Case A. The similar conclusion also holds for the
$h\rightarrow VV$ mode.
On the other hand, $\gamma\gamma$ and $VV$ decay modes of Higgs boson produced
via weak boson fusion will be quite different from that via
$gg$ fusion since the weak boson fusion process is not largely suppressed.
In Case A, $R_{\sigma(VV)}\times R_{{\rm BR}(\gamma \gamma)}$ is very
close to the SM prediction. In Case B, however, it could be significantly
enhanced because of the enhancement of the $\gamma\gamma$ decay 
branching ratio.
The decay branching ratio to $\tau\tau$ decay mode and the Higgs boson
production
rate via weak boson fusion do not change in either Case A or Case B.
Thus, $R_{\sigma(VV)}\times R_{{\rm BR}(\tau \tau)}$ is close
to the SM prediction.

When Higgs mass is relatively heavy ($m_h>160$ GeV),
the decay mode to gauge bosons ($h\rightarrow VV$ ($V=Z,W$)) becomes
important. Since in both Case A and Case B the branching ratio for 
$h\rightarrow VV$ ($V=Z,W$) is almost the same as the SM prediction, 
the $VV$ decay mode via gluon fusion production is significantly
suppressed, but that via weak boson fusion is not.
Therefore, typically the discovery modes of Higgs boson produced 
via weak boson fusion processes will become 
more important in
the LH model than in the SM.
Since these effects could be larger than the experimental 
uncertainties $(10\%-20\%)$~\cite{Zeppenfeld:2000td}
on the measurement of the Higgs boson
production cross sections times the branching ratios,
they could be observable at the LHC.\footnote{We stress that
the further improvement of the theoretical calculation will
be important to observe these effects.}

We note that when the scale $f$ is around 1 TeV, the T-odd heavy gauge boson
masses can be of $O(100)$ GeV. For example, for $f=700$ GeV,
the T-odd $U(1)$ ($A'$) and $SU(2)$ ($W'$ and $Z'$) gauge boson
masses are $m_{A'}\simeq 100$ GeV and $m_{W',Z'}\simeq 450$ GeV,
respectively.
When the Higgs mass is
larger than twice of these masses, Higgs boson can decay into
the heavy gauge boson pair. We have checked that in that case
the decay branching ratios
of the heavy gauge boson pair
are less than $10^{-2}$ for $f\geq 700$ GeV. Therefore, the extra Higgs decay modes
does not significantly change the branching ratios of the SM Higgs 
boson decay modes.

As discussed in previous sections, 
the Higgs boson production rate via $gg\rightarrow h$ process
in the Littlest Higgs model with T-parity strongly depends
on the mechanism to give T-odd particles (and other additional extra-particles)
heavy masses. Thus the prediction may be sensitive to new 
physics above the cutoff scale, and our result shows that the effect could be
quite large and become observable at the LHC.
Therefore, searching for Higgs boson in various detection modes at the LHC is
very important, and the measurement of the relative event rates
in multi-channels could reveal the mechanism which provides
the cancellation of the quadratic divergence of the Higgs mass
parameter and the origin of mass terms for the extra heavy fermions
in the LH model.

\section*{Acknowledgments}
We thank A. Belyaev, R.S. Chivukula and Jay Hubisz for helpful 
discussions.
This work was supported in part by the U. S. National Science
Foundation under award PHY-0244919.
The Feynman diagrams in Fig.~\ref{ggh} was drawn with 
{\tt JaxoDraw}~\cite{Binosi:2003yf}.

\end{document}